\documentclass[%
 reprint,
superscriptaddress,
 amsmath,amssymb,
 aps,
 prl,
]{revtex4-1}

\usepackage{textcomp}

\usepackage{graphicx}% Include figure files
\usepackage{epstopdf}
\usepackage{xcolor}
\usepackage{dcolumn}% Align table columns on decimal point
\usepackage{bm}% bold math
\usepackage{hyperref}% add hypertext capabilities
\hypersetup{colorlinks,linkcolor=blue,citecolor=blue}
\begin{document}

\preprint{APS/123-QED}

\title{An atom interferometer driven by a picosecond frequency comb}

% Force line breaks with \\

\author{Cyrille Solaro}
\email{cyrille.solaro@lkb.upmc.fr}
\author{Cl\'{e}ment Debavelaere}
\author{Pierre Clad\'{e}}
\affiliation{Laboratoire Kastler Brossel, Sorbonne Universit{\'{e}}, CNRS, ENS-Universit{\'{e}} PSL, Coll{\`{e}}ge de France, 75005 Paris, France}
\author{Sa\"{i}da Guellati-Khelifa}
\email{guellati@lkb.upmc.fr}
%\email{cyrille.solaro@lkb.upmc.fr}
\affiliation{Laboratoire Kastler Brossel, Sorbonne Universit{\'{e}}, CNRS, ENS-Universit{\'{e}} PSL, Coll{\`{e}}ge de France, 75005 Paris, France}
\affiliation{Conservatoire National des Arts et M\'{e}tiers, 75003 Paris, France}

\date{\today}% It is always \today, today,

\begin{abstract}
We demonstrate a light-pulse atom interferometer based on the diffraction of free-falling atoms by a picosecond frequency-comb laser. More specifically, we coherently split and recombine wave packets of cold $^{87}$Rb atoms by driving stimulated Raman transitions between the $|5s~^2S_{1/2},F=1\rangle$ and $|5s~^2S_{1/2},F=2\rangle$ hyperfine states, using two trains of picosecond pulses in a counter-propagating geometry. We study the impact of the pulses length as well as of the interrogation time onto the contrast of the atom interferometer. Our experimental data are well reproduced by a numerical simulation based on an effective coupling which depends on the overlap between the pulses and the atomic cloud. These results pave the way for extending light-pulse interferometry to transitions in other spectral regions and therefore to other species, for new possibilities in metrology, sensing of gravito-inertial effects and tests of fundamental physics.  
\end{abstract}

\maketitle

%\cite{Holzwarth2000,Jones2000} % invention Frequency combs
%\cite{Haensch2006,Hall2006} % Nobel lectures
%\cite{Marian_2004} % visible Rubidium single/two photon few 10kHz Jun Ye
%\cite{Wolf2009} % visible one photon 393nm Calcium 500 kHz unpconverted crystal Eikema
%\cite{Eckstein1978} % visible two photon Na 500 kHz Hansch
%\cite{Barmes2013} % visible two photon Rb Cs 2x760 et 2x822 nm Eikema
%\cite{Cingoez2012} % HHG XUV one photon Ag 86nm 3MHz Jun Ye
%\cite{Witte2005} % deepUV two photon 2x212 nm Kr Eikema
%\cite{Kandula2010} % XUV Pair of pulses He4 ionization energy 6MHz 51nm Eikema 
%\cite{Yost2016} % deepUV two photon 1S-3S 2x205nm hydrogen crystals Hansch Udem
%\cite{Altmann_2016} % deepUV two photon 2x212 nm Kr crystals Eikema 
%\cite{Altmann2018} % deepUV two photon  H2 2x201 nm 70kHz Eikema
%\cite{Dreissen2019} % XUV two photon 2x110 nm Xe Eikema 
%\cite{Grinin2020} % deepUV two photon 1S-3S 2x205nm hydrogen crystals Hansch Udem
%
%\cite{Porat2018} % HHG XUV challenge

An optical frequency comb is a laser source whose spectrum consists of a series of phase-coherent evenly spaced narrow frequency lines. In the time domain, it is a source of precisely timed phase-coherent ultrashort pulses with high peak intensities, particularly suitable for frequency conversion processes in non-linear media. For this reason, frequency combs are very promising for precision measurements in spectral regions that are not easily accessible with continuous-wave (CW) lasers (e.g. deep-, vacuum- or extreme-ultraviolet (XUV)). 
For instance, laser frequency combs have already been used for high-resolution spectroscopy \cite{Picque2019} in the deep-UV \cite{Witte2005,Yost2016,Altmann_2016,Altmann2018}, vacuum-UV \cite{Dreissen2019} and XUV \cite{Kandula2010,Cingoez2012}. In particular, in Ref. \cite{Grinin2020}, Grinin et \textit{al.} probed with sub-kHz accuracy a two-photon transition (at $2\times205$ nm) in atomic hydrogen with a picosecond frequency comb, out-passing the state-of-the-art achieved with CW lasers and demonstrating the huge potential of frequency combs for high precision spectroscopy in the deep-UV. 
In this context, it is very worth exploring the potential of frequency combs for light-pulse atom interferometry as well. In fact, light-pulse atom interferometry, where light pulses are used as atom beam splitters, has so far only exploited CW laser sources and is thus limited to a handful of atomic species.
Yet, extending this technique to other spectral regions could open-up new possibilities for tests of fundamental physics. Especially, it could be used on anti-hydrogen in the deep-UV to measure the free-fall acceleration of anti-matter $\bar{g}$. This would allow for an interferometry test of the gravitational interaction between matter and anti-matter that would be orders of magnitude more stringent than the classical tests currently underway at CERN \cite{Bertsche2018,Mansoulie2019,Pagano2020}, which aim relative accuracies on $\bar{g}$ of $\sim 10^{-3}$ at best. 

In this Letter, we report on a light-pulse atom interferometer driven by a picosecond frequency comb. We demonstrate this technique in the visible spectrum on a free-falling cloud of laser-cooled $^{87}$Rb atoms, in a configuration where the interferometer is sensitive to the Earth gravitational acceleration $g$. We discuss the different mechanisms at play and show that a Monte Carlo simulation based on an effective coupling for the atom-comb interaction reproduces well our experimental data.

\begin{figure*}[ht]
\includegraphics[width=1\textwidth]{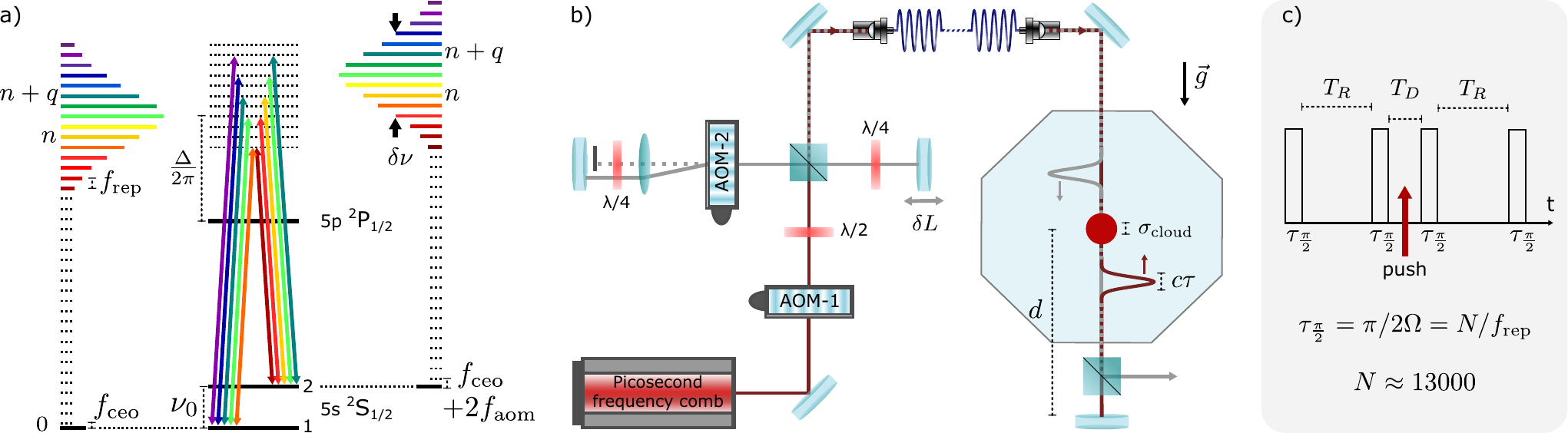}
\caption{a) Principle of frequency-comb-driven Raman transitions between two hyperfine levels in the electronic ground state of $^{87}$Rb. b) Schematic overview of our experimental setup. The overlap position of the counter-propagating pulses is adjusted precisely by translating one of the mirrors in the delay line. The Doppler effect due to free fall is compensated by chirping the frequency of an acousto-optic modulator (AOM-2). c) Pulse sequence of the Ramsey-Bord\'{e} interferometer. Each $\pi/2$-pulse acts as an atom beam splitter and is realized by a train of $N \approx 13000$ picosecond pulses. }
\label{Fig_Exp}
\end{figure*}

The atom beam splitters are realized by driving, with the frequency comb, stimulated Raman transitions between the $|F=1\rangle$ and $|F=2\rangle$ hyperfine levels of the $5s~^2S_{1/2}$ electronic ground state. The principle of frequency-comb-driven Raman transitions can be understood in the frequency domain (see Fig.\ref{Fig_Exp}.a)) as a coherent sum of many stimulated Raman processes, each induced by pairs of comb lines which frequency difference, a multiple of the comb's repetition rate $f_\text{rep}$, verifies $qf_\text{rep}=\nu_0$, where $\nu_0$ is the Raman transition frequency (here $\nu_0\sim 6.8$ GHz). If the comb's spectral bandwidth $\delta\nu \gtrsim \nu_0$, all frequency components can contribute to drive the transition \cite{Solaro_2018}. 
In the time domain, frequency-comb-driven Raman transitions can be understood as the constructive interference of state coherences that are periodically induced by the train of ultrashort pulses (with period $T_\text{rep}=1/f_\text{rep}$). Such coherences are induced efficiently if the pulse duration $\tau$ is short compared to their free evolution $T_0\equiv 1/\nu_0 \gtrsim \tau$ (condition of impact excitation \cite{Fukuda1981}, equivalent to $\delta \nu \gtrsim \nu_0$) and interfere constructively if $T_\text{rep}=qT_0$. 
Previously, stimulated Raman transitions had been driven with frequency combs in atomic vapors \cite{Fukuda1981} and more recently in trapped atomic and molecular ions for quantum information processing \cite{Hayes2010,Lin2020} and high-resolution spectroscopy \cite{Solaro_2018,Solaro_2020,Chou2020}.

In contrast with frequency-comb-driven Raman spectroscopy experiments, it is crucial for atom interferometry that the stimulated Raman process be driven by two trains of ultrashort pulses in a counter-propagating geometry. This enables the spatial separation of the two interfering atom wave packets thanks to the transfer of two photon recoils. In this configuration, each pair of counter-propagating ultrashort pulses must overlap at the atoms position. First, this imposes that the pulse length be on the order of the cloud size ($c\tau\sim\sigma_\text{cloud}\sim$ \mbox{1 mm}). To verify both this condition and the condition of impact excitation for typical hyperfine splittings, one must use picosecond pulses. Second, this requires a good control of the pulses overlap position and, in the case of free-falling atoms, this limits the interferometer duration to $\sqrt{2c\tau/g}\sim 10$ ms for a picosecond pulse of \mbox{$\tau\sim 2$ ps}. 
Finally, as for traditional interferometers using CW lasers \cite{Kasevich_Chu_1991,Giltner_&al_1995}, it is also necessary to control the phase difference of the counter-propagating pulses precisely and to compensate for the Doppler effect during free-fall.

These qualitative arguments can be obtained formally by calculating the Raman transfer probability after $N$ pairs of counter-propagating picosecond pulses. To do so, we can first eliminate the excited $5p~^2P_{1/2}$ state (see figure \ref{Fig_Exp}.a)) adiabatically, and keep only the two ground states $|5s~^2S_{1/2},F=1\rangle$ and $|5s~^2S_{1/2},F=2\rangle$ with an effective 2-photon coupling. We then use an appropriate rotating frame where the system's Hamiltonian is periodic (of period $T_\text{rep}$) so that after a time $t=NT_\text{rep}$, the associated evolution operator verifies $U_\text{rot}(NT_\text{rep})=U_\text{rot}(T_\text{rep})^N$. One can then show that:
\begin{equation}\label{Eq_Urot}
U_\text{rot}(t)=
e^{-\frac{i}{\hbar}\left[\begin{matrix}0 & \frac{\Omega(z, \tau)}{2} \\
\frac{\Omega(z, \tau)}{2} & \delta\end{matrix}\right]t}
\end{equation}
where $\delta/2\pi=\nu_0-\nu-qf_\text{rep}$ (with $\nu$ the difference in carrier frequency of the two counter-propagating pulses) and 
\begin{equation}\label{Eq_Omega_eff}
\Omega(z, \tau) = \frac{\Gamma^2}{4\Delta}\frac{I}{2I_s}e^{-(\frac{z}{c\sigma_\tau})^2} e^{-(\frac{(\omega_0-\omega)\sigma_\tau}{2})^2}
\end{equation}
where $\Gamma$, $\Delta$ and $I_s$ are the natural linewidth, the 1-photon detuning and the saturation intensity of the $5s~^2S_{1/2} - 5p~^2P_{1/2}$ transition, respectively. Here, $I=I_0\sqrt{\pi}\sigma_\tau f_\text{rep}$ is the combined average intensity of the two trains of picosecond pulses, which intensity envelops are here considered to be Gaussian (with peak intensity $I_0/2$). Eq.\ref{Eq_Urot} is valid in the limit $\Omega T_\text{rep}\ll 2\pi$, or equivalently when a Raman transfer probability of unity requires a large number of picosecond pulses ($N\gg 1$).  The atom-comb interaction can thus be seen as an effective interaction between a two level atom and a CW-laser, with an effective detuning $\delta$ and an effective coupling strength $\Omega(z, \tau)$, which quantify each of the three conditions discussed above. $\delta$ originates from the accumulated dephasing of each periodic excitations induced by the picosecond pulses with respect to the Raman transition. The spatial dependence of $\Omega(z,\tau)$ is deduced from the instantaneous effective two-photon coupling. The impact condition, given by the second exponential term in Eq.(\ref{Eq_Omega_eff}), results from the Fourier transform of the pulse shape when integrating the Hamiltonian for one pulse.

%Our experimental cycle lasts $\sim 1.5$ s and consists in a preparation sequence, an interferometer sequence and a detection sequence. The experimental starts with a preparation sequence during which we produce a cloud of $\sim10^8$ $^{87}$Rb atoms at a temperature of $\sim6$ \textmu{}K using standard laser cooling and trapping techniques.

An overview of our experimental setup is presented in Fig.\ref{Fig_Exp}.b). 
The frequency comb is a picosecond mode-locked Ti:sapphire solid-state laser from Coherent Inc. (model Mira 900-P \cite{mira}). Its repetition rate \mbox{($f_\text{rep}\sim 76$ MHz)} is phase-locked to a synthesizer referenced to a cesium frequency standard by a servo-loop controlling the position of one of the cavity mirrors with one fast and short-range piezo-electric element and one slow and long-range piezo-stack. 
This servo-loop does not eliminate phase noise at 100 Hz which we attribute to intensity noise of the pump laser. To mitigate this noise, we synchronize the experiment with the AC wall voltage (at 50 Hz).
The pulse duration can be slightly varied from $\tau\sim 1$ to 2 ps (FWHM of intensity envelope) by adjusting the width of a slit in the laser cavity. The corresponding spectral bandwidth ($\delta\nu\sim 160$ to 320 GHz) is measured continuously with a spectrometer (resolution of $\sim 20$ GHz) previously calibrated. %(HighFinesse \AA ngstom WS-U)
An auto-correlation measurement using an acousto-optic modulator as non-linear medium was also performed at \mbox{$\tau = 2$ ps} to verify that the pulse is Fourier-transform-limited. The detuning from the $5s~^2S_{1/2} - 5p~^2P_{1/2}$ transition is set to \mbox{$\Delta/2\pi=0.41$ THz} (corresponding to a laser wavelength of \mbox{$\lambda = 794.1$ nm}) by rotating the birefringent filter in the cavity. 

The optical setup used to control precisely the overlap position of the counter-propagating pulses and to compensate for the Doppler effect is illustrated figure \ref{Fig_Exp}.b). A first acousto-optic modulator (AOM-1) is used as a switch to generate the trains of picosecond pulses. Each picosecond pulse is split in two by a polarizing beam splitter. %A half-wave plate is used to balance the optical power in the two arms.
The transmitted pulse (in brown on Fig.\ref{Fig_Exp}.b)) is sent directly to the vacuum chamber via an optical fiber. After propagating through the chamber, it is retro-reflected by a horizontal mirror placed below the chamber at an optical distance of $d \sim 37$ cm from the atoms position. The optical path in the chamber is aligned with the vertical axis. 
The reflected pulse (in gray on Fig.\ref{Fig_Exp}.b)) enters an adjustable delay line (with an optical length equal to $2d$) prior to be coupled in the same optical fiber. The overlap position of the two counter-propagating pulses is precisely adjusted by controlling the length of the delay line. %which is placed on a translation stage. 
The laser beams $1/e^2$ radii are \mbox{$w_0=2.0$ mm} at the atoms position. With an average power of $20$ mW per beam, the effective Raman Rabi frequency is \mbox{$\Omega/2\pi \approx 1.4$ kHz} and the $\pi$-pulse duration ($\tau_\pi \approx 350$ \textmu{}s) corresponds to a train of $\tau_\pi f_\text{rep}\approx 26000$ picosecond pulses. The reflected pulse (gray) is shifted in frequency by $\nu=2f_\text{AOM}$ with a second AOM used in a cat's eye configuration. Chirping the AOM frequency linearly during the interferometer ($2f_\text{AOM}(t)=2k\alpha t$ where $k=2\pi/\lambda$) allows to compensate for the Doppler effect due to free fall. 

\begin{figure}[t]
\includegraphics[width=1\columnwidth]{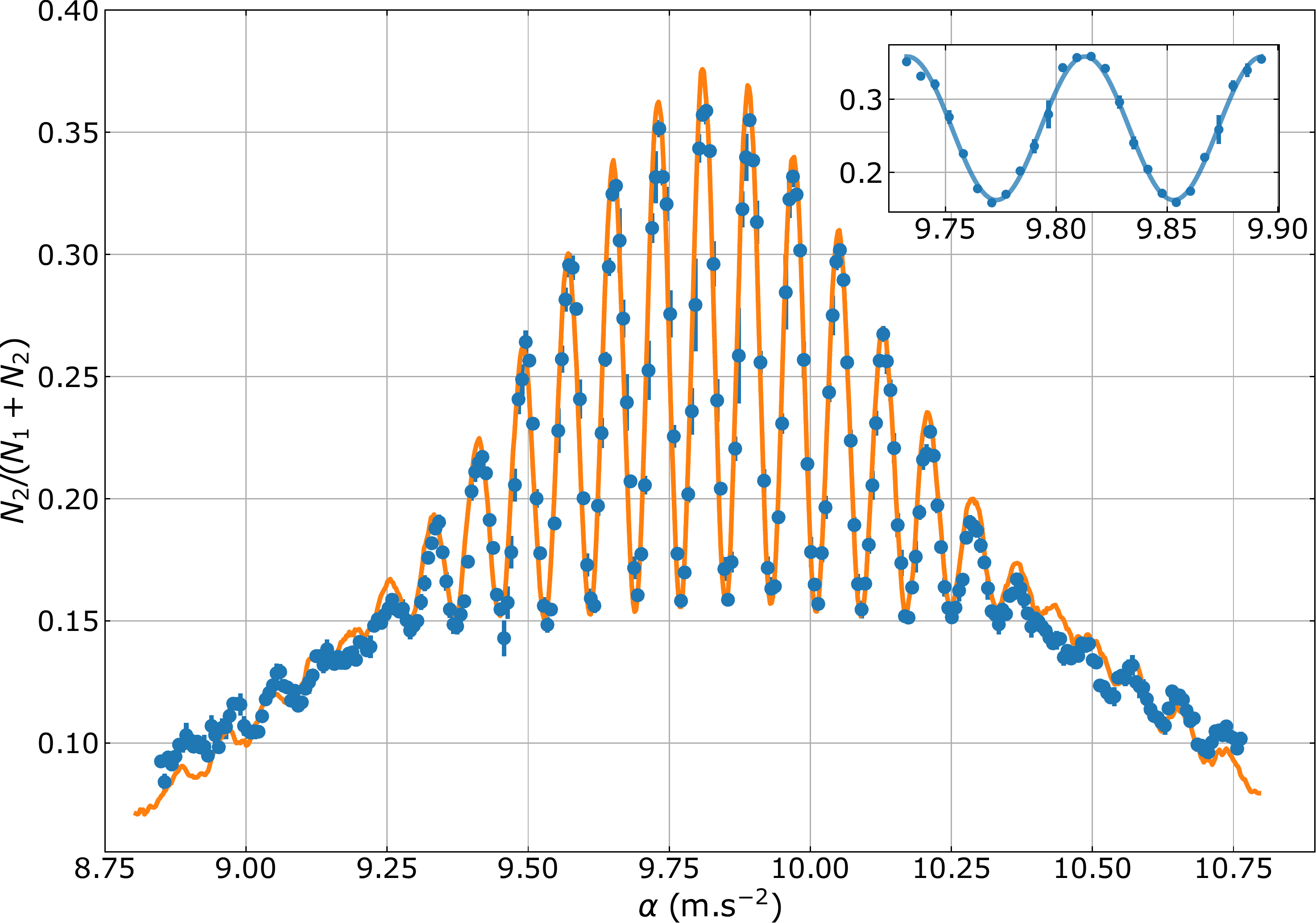}
\caption{Atomic fringes at the output of the interferometer: transfer probability to the $|5s~^2S_{1/2},F=2\rangle$ state as a function of the frequency chirp used to compensate for the Doppler effect due to free fall. Experimental data are shown as blue points. The simulated fringes without any adjusted parameters are shown as a solid orange line (see text for more details). Here $\Delta/2\pi=0.41(5)$ THz, $T_R=1.5$ ms, $T_D=1.2$ ms, $\tau=2.00$ ps and $\sigma_\text{cloud}=0.67$ mm. Inset zoom: A fit (solid line) to the central fringe is used to determine the fringe contrast and the Earth gravitational acceleration with a relative statistical uncertainty $\sim 10^{-5}$.}
\label{Fig_Fringes}
\end{figure}

We implement a so-called Ramsey-Bord\'{e} interferometer made of two pairs of atom beam splitters (i.e. $\pi/2$-pulses) as illustrated in figure \ref{Fig_Exp}.c). Here, each atom beam splitter is realized by two counter-propagating trains of $N\approx 13000$ picosecond pulses. After the first pair of atom beam splitters, a CW-laser pulse resonant with the $|5s~^2S_{1/2},F=2\rangle - |5p~^2P_{3/2},F=3\rangle$ transition pushes away atoms in the $|F=2\rangle$ state to increase the contrast of the interferometer. Following this interferometer sequence, the number of atoms in each hyperfine state $|F=1\rangle$ and $|F=2\rangle$ is measured by state selective fluorescence detection. Atomic fringes are obtained by recording the fraction of atoms in a given state for varying chirp parameter $\alpha$. A typical set of data obtained at the output of the interferometer is presented in figure \ref{Fig_Fringes}. Each data point corresponds to an average over three repetitions of the experimental cycle. The inset in figure \ref{Fig_Fringes} shows a fit of the central fringe from which we deduce the contrast of the interferometer. The uncertainty on the central fringe position allows for a determination of the Earth gravitational acceleration $g$ with a relative uncertainty on the order of $10^{-5}$.

To model the interferometer, we implemented a Monte Carlo simulation based on equations Eq.(\ref{Eq_Omega_eff}). The initial positions and velocities of $10^6$ atom are randomly chosen according to the initial distribution of positions and velocities in the cold atom cloud. The classical trajectories of each atom during free-fall is calculated and the amplitudes of the two-states superposition evaluated after each of the four $\pi/2$-pulses using Eq.(\ref{Eq_Omega_eff}). We obtain the proportion of atoms in each of the two hyperfine states at the end of the atom interferometer. According to the simulation, the total fraction of atoms participating to the interferometer is about 15 \textperthousand{} for our experimental parameters. 
This value is in good agreement with our measurements. It is limited by the duration of the picosecond pulse: varying $\tau$ from e.g. 1 to 2 ps increases the pulses overlap volume by a factor two and, if $\sigma_\text{cloud}>c\tau$ as in our experiments, it increases the fraction of atoms participating to the interferometer by the same factor. 
%With a cloud temperature of $\sim 6$ \textmu{}K, the Doppler broadening is $\sim$ kHz (at FWHM), a factor of ten larger than the effective Rabi frequency and
It is also limited by the cloud temperature ($\sim 6$ \textmu{}K) which results in a Doppler broadening of a few tens of kHz that is larger than the effective Rabi frequency.
To simulate the atomic fringes, we evaluate the proportion of atoms in each hyperfine states for varying chirp parameter $\alpha$ (orange line in figure \ref{Fig_Fringes}). The simulated interferometer contrast is deduced from a fit of the simulated central fringe and compared to our experimental data.

\begin{figure}[h]
\includegraphics[width=1\columnwidth]{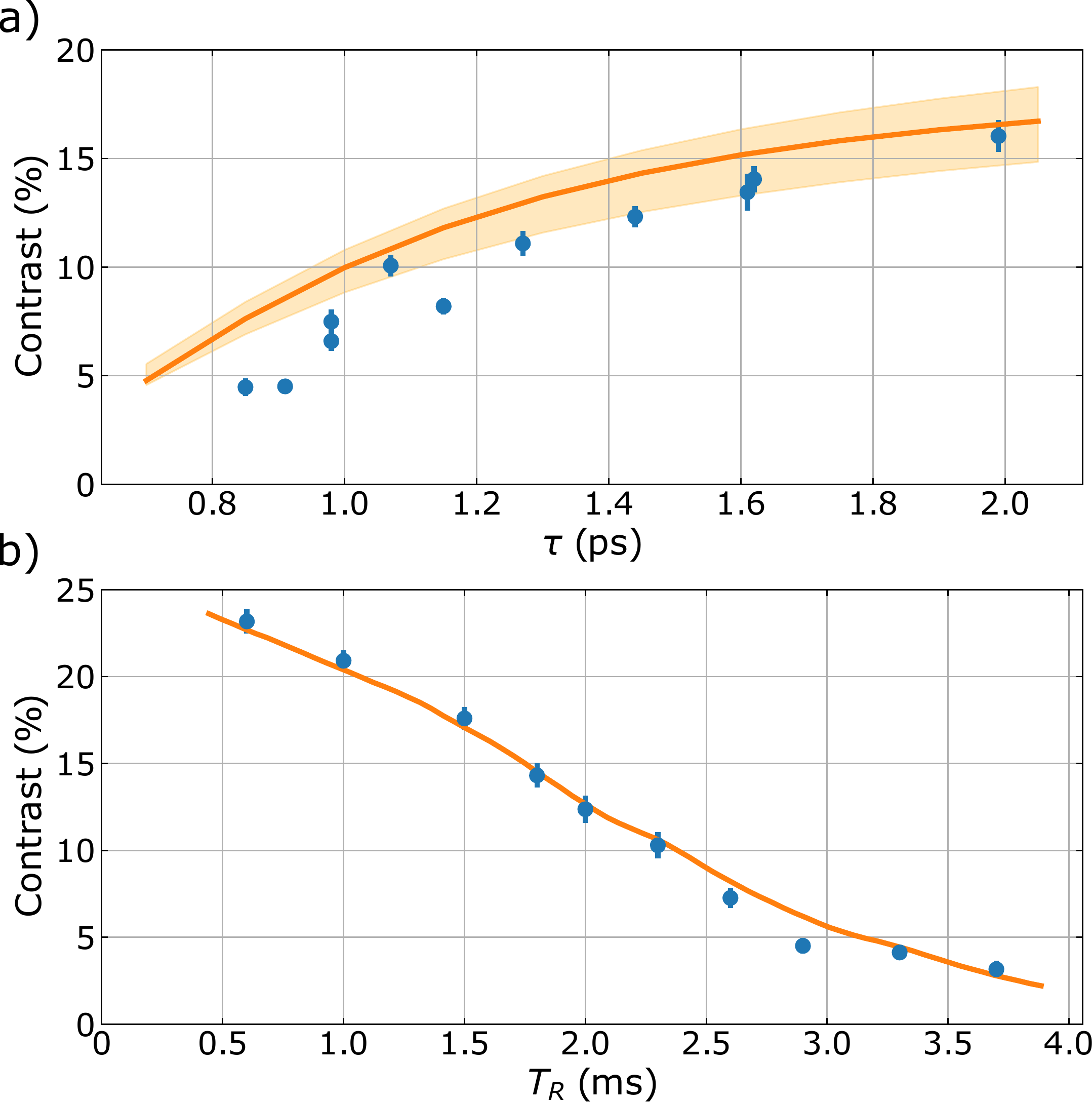}
\caption{a) Contrast of the central fringe as a function of the picosecond pulse duration, for $\sigma_\text{cloud}=0.97$ mm, $\Delta/2\pi=0.87(5)$ THz, $T_R=1$ ms and $T_D=1.2$ ms. Experimental data are shown as blue points. The simulation without adjusting parameters is shown as a solid line, while the shaded area correspond to the simulated 1$\sigma$ uncertainty due to the uncertainties on the experimental parameters showing worst and best case scenarios. For short pulse duration, the discrepancy is explained by the increase of spontaneous emission due to a larger comb bandwidth. b) Contrast of the central fringe as a function of $T_R$ with $T_D=1.2$ ms. Here, $\tau=2.00$ ps, $\sigma_\text{cloud}=0.67$ mm and $\Delta/2\pi=0.51(5)$ THz. %The experimental data (blue points) are well reproduced by our simulation (orange points) without any adjusting parameters.
}
\label{Fig_Contrast}
\end{figure}

As shown in figure \ref{Fig_Contrast}, our simulation results reproduce our experimental data with good quantitative agreement. 
The solid curves in the two graphs correspond to the simulated contrast without any adjusted parameters. The orange shaded areas represent the best and worst case scenarios deduced from the uncertainties in the measured parameters, mainly $\Delta$, $I$ and $\sigma_\text{cloud}$. Figure \ref{Fig_Contrast}.a) shows the impact of the picosecond pulses duration on the contrast of the interferometer. As expected, increasing the pulse duration increases the contrast of the interferometer since the atoms, subject to gravity, are freely falling through the pulses overlap volume during the interferometer sequence. For short picosecond pulses, the discrepancy is explained by an increase of spontaneous emission due to the larger comb bandwidth, as observed in the experiment. 
Figure \ref{Fig_Contrast}.b) presents the evolution of the interferometer contrast as a function of the interrogation time $T_R$, for a pulse duration of \mbox{$\tau=2.00$ ps}. The contrast decays as the interrogation time is increased and is almost zero for \mbox{$T_R\sim4$ ms}. This behavior is well reproduced by our simulation, it is a consequence of the finite volume of the pulses overlap. In fact, for \mbox{$T_R\sim 4$ ms}, the total interferometer time is \mbox{$\sim 10$ ms} and corresponds to a free fall distance of \mbox{$\sim 1$ mm} on the order of the pulse length $c\tau$. This limitation is not fundamental. Longer interaction times, and therefore better sensitivities on $g$, could be achieved by adjusting the delay line during the atom interferometer so that the position of the pulses overlap volume follows the atom trajectory during free-fall.

In conclusion, we have demonstrated a frequency-comb-driven atom interferometer based on the diffraction of atoms by two counter-propagating trains of picosecond pulses. We studied the impact of both the picosecond pulses duration and the interrogation time on the contrast of the interferometer in a vertical geometry where the atoms are subject to gravity. The experimental data are reproduced with good quantitative agreement by a Monte-Carlo simulation based on an effective coupling which depends on the atoms positions with respect to the pulses overlap region. % reproduces the experimental data with good quantitative agreement. 
Although we implemented it in the visible spectrum around 800 nm to drive Raman transitions in rubidium atoms, this technique holds great promises for extending atom interferometry to other spectral regions (deep-UV, V-UV, X-UV). Indeed, one can benefit from the high peak intensity of the ultrashort pulses which makes frequency conversion in crystals and gas targets more efficient, as shown by the remarkable progress in the generation of XUV frequency combs for high precision spectroscopy \cite{Nauta2017,Porat2018,Dreissen2019,Nauta2021}.
Therefore, frequency-comb-driven atom interferometry could eventually open the door for extending interferometry to new transitions and new species for novel possibilities in metrology, sensing of gravito-inertial effects and tests of fundamental physics. The modest relative sensitivity ($\sim 10^{-5}$) on the Earth gravitational acceleration we demonstrated with this technique, if it were to be reproduced on the $1s~^2S_{1/2} - 2p~^2P_{1/2}$ transition at 121 nm in anti-hydrogen, would lead to a stringent test of the weak equivalence principle with anti-matter.

\begin{acknowledgments}

We are extremely grateful to F. Nez for lending us the MIRA-900-P frequency-comb laser and we thank C. Carrez for his contribution in the early stage of the experiment. C.S. acknowledges support from Region Ile-de-France through the DIM SIRTEQ Fellowship ELUDA, and from the LabEx ENS-ICFP: ANR-10-LABX-0010/ANR-10-IDEX-0001-02 PSL*.

\end{acknowledgments}

\bibliography{Everything}

\end{document}